# Discussing Privacy and Surveillance on Twitter: A Case Study of COVID-19
Jayati Dev

**Introduction**

Technology is uniquely positioned to help us analyze large amounts of information to provide valuable insight during widespread public health concerns, like the ongoing COVID-19 pandemic. In fact, information technology companies like Apple and Google have recently launched tools for contact tracing - the ability to process location data to determine the people who have been in contact with a possible patient, in order to contain the spread of the virus [1]. While China and Singapore have successfully led the effort [2], more and more countries are now implementing such surveillance systems, raising potential privacy concerns about this long term surveillance. For example, it is not clear what happens to the information post-pandemic because people are more likely to share their information during a global crisis without governments having to elaborate their data policies [3]. Digital Ethnography on Twitter, which has over 330 million users worldwide, with a majority in the United States where the pandemic has the worst effects [4] provides a unique opportunity to learn about real-time opinions of the general public about current affairs in a rather naturalistic setting. Consequently, it might be useful to highlight privacy concerns of users [6], should they exist, through analysis of Twitter data and information sharing policies during unprecedented public health outbreaks. This will allow governments to protect its citizens both during and after health emergencies.

The specific research questions in this study would be to see how the discussion around privacy and surveillance has evolved over the duration of the Covid-19 pandemic. They are as follows:
1. What are the various discussion topics involving Covid-19 surveillance and how frequently do they occur?
2. What are users' sentiments about surveillance during the Covid-19 outbreak?

Using Python libraries for topic modelling using Latent Dirichlet Analysis (LDA) and sentiment analysis using Natural Language Toolkit (NLTK), I report the discussions around privacy and people's sentiments towards surveillance at large. I also observe the discussion over time and user engagement on Twitter in these topics, which reveal that users engage in privacy discussions around COVID-19 possibly propelled by popular media articles with a rising negative sentiment for government surveillance (and other privacy concerns). The findings indicate a need for better planning in data collection and analysis by governments and companies that are privacy-preserving while providing an important information source for governments in containing public health emergencies like COVID-19.

**Related Work**
The current COVID-19 pandemic has raised important questions about the way we deal with privacy and security concerns of personal information in the wake of a public health emergency. A number of countries have implemented widespread surveillance of its citizens by using location information for contact-tracing. This helps them understand if people with COVID-19 symptoms have been in contact with other people who in turn might get infected. While articles claim that [3] surveillance has not been particularly effective in controlling the outbreak, many countries argue otherwise [11]. This has raised concerns among privacy think- tanks about what would be done with user information, which users readily provide in efforts to contain the pandemic, after the outbreak [3]. With Google and Apple combining efforts for contact-based tracing using granular location information [1], people who are vulnerable during the outbreak, must not be affected by the consequences of Big Data collection after the same.

Furthermore, the extent of remote work and school during the novel coronavirus outbreak has also enabled people to connect with their workplace and academic work from home via the Internet while maintaining social distancing norms. Video conferencing software such as Zoom have been found to collect user information and have multiple security bugs that lead to 'Zoom-bombing' [12], with several efforts being made by Zoom to fix such bugs. This also creates a need to study the privacy concerns and sentiment around technological problems (or misuse) of products that are instrumental for enabling people to be connected during extended periods of social isolation and government lockdowns.

Though Twitter is often studied for user sentiment in socio-political contexts like the spread of false information [9], it is rarely used for evaluating privacy concerns among people. Privacy concerns are usually studied through traditional mixed method research tools like surveys and interviews. To the best of my knowledge, there is limited research on privacy concerns during such a large scale emergency like never before. However, despite its limitations, Twitter provides a naturalistic setting to understand popular conversation about privacy and security that emerge as an indirect effect of the novel coronavirus outbreak. The variety of discussion on Twitter provides a starting point for a more nuanced analysis of privacy concerns and has been the focus of this study.

**Method**
While Twitter users tend to be younger, more educated, and more liberal than the general population [5], it nevertheless provides an opportunity to provide insight into privacy concerns of individuals. The data was collected using Get Old Tweets API[1] that maintains an archive of old tweets. Newer tweets (the last 100 tweets in the dataset) have been collected using the Tweepy Twitter API[2]. I conducted a time-series analysis from March 1, 2020 (first week of government lockdown [7]) to April 20, 2020 (current date of data collection) to measure the number of tweets by users over time. This was followed by a measure of occurrences of retweets and favorites for the specific tweets collected to study how users engaged with privacy-specific content regarding COVID-19 on Twitter. I used a predetermined set of keywords that include "coronavirus" and "privacy".

---

[1] https://github.com/Jefferson-Henrique/GetOldTweets-python

The first research question required an in-depth analysis of tweets since this is a novel phenomenon with limited previous research theories. I used Latent Dirichlet Analysis for topic modelling in order to highlight the different privacy themes and opinions that emerge. This was done using the LDA model available through the `gensim` package in Python and the Natural Language Toolkit (NLTK). NLTK was used to tokenize words from tweets, remove existing stop words (like prepositions), and reduce the words to its stem form. The number of topics in LDA was set to 10 for the dataset (tweets from March and those from April respectively). The resulting topics were then tagged with a description to address the emerging topic.

In order to answer the second research question, I performed a sentiment analysis of tweets using Python's using a naive Bayes classification approach (previously known as automatic indexing [8]). I used `pandas` for string handling and converting the dataset from `csv` into a `pandas` dataframe for easy manipulation. I also used `sklearn` to access the Bayes classification algorithm and NLTK packages for data pre-processing. `Sklearn` was also used to create datasets for both training and testing the sentiment analysis model (I used the same dataset for both). The results from the analysis are described below. Please refer to the appendix for code and datasets.

**Ethics:** I chose not to collect Twitter user information in order to make the dataset de-identified and protect user privacy. The collection of publicly available tweets does not require an Institutional Review Board (IRB) application.

**Findings**

The resulting dataset from mining Twitter for March-April 2020 generated a total of 22208 tweets (April - 10567; March - 11641). I used time-series analysis for tweets, retweets, and favorites, to see the number of discussions over time and the impact per discussion (tweet) during the pandemic. This is followed by topic modelling for the 10 topics that people talk about regarding privacy specific to surveillance during Covid-19, and a sentiment analysis of such tweets (positive, negative, or neutral). The findings are explained in detail below.

---

[2] https://www.tweepy.org/ For simplicity, I have uploaded the dataset from Get Old Tweets to have everything in a single dataset.

**Privacy-Related Tweets Decrease over Time**

Figure 1 shows the number of tweets over time since March 1, 2020 that express surveillance concerns. The graph peaks on March 23, 2020, which is the date of publication of the first Electronic Frontier Foundation article on government surveillance during COVID-19 and the resulting privacy concerns. The graph first sees a spike during the first week of lockdown (which started on March 13, 2020 in the United States). A polynomial trendline ($R^2$=0.583) shows that tweets expressing privacy concerns peaked during the first week of April and then subsequently started to decrease in frequency.

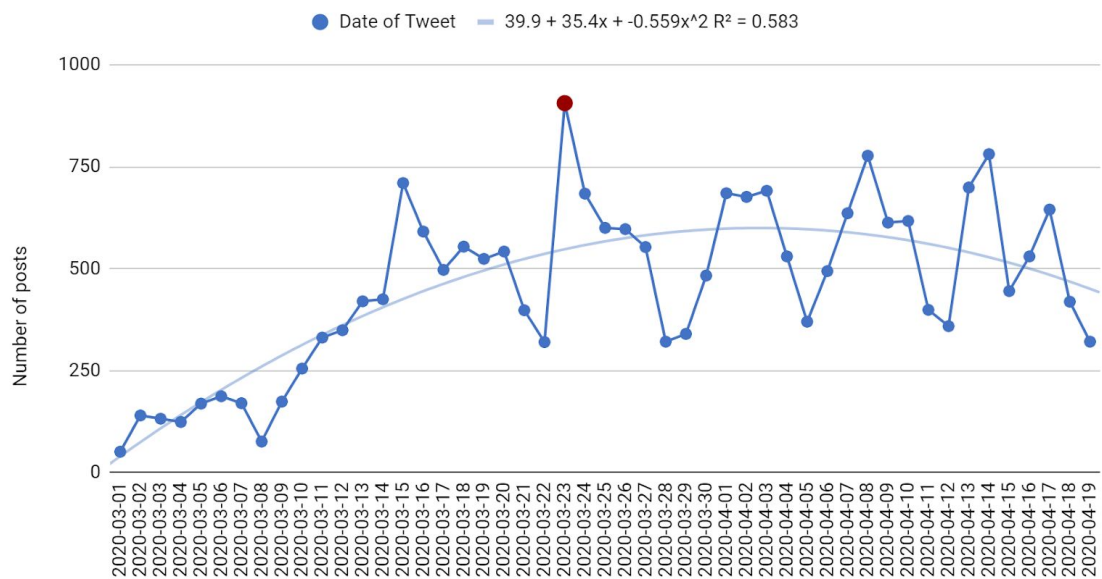

Figure 1: Tweets per day during March 1 - April 20, 2020 with the keywords "privacy" and "COVID-19" or related.

**User Engagement is Disproportionate**

Analysis of the retweets and favorites for each tweet presented an interesting result. As can be seen in Figure 2 and Figure 3, There is a disproportionate amount of engagement with specific tweets while most tweets usually have a number of retweets and favorites under the numeric 10 baseline. Table 1 shows the mean, median, maximum value, and minimum value of retweets and favorites respectively for all the 22208 tweets. As mentioned in the table, I had to remove an outlier tweet (with retweets = 7743 and favorites = 16943 as of April 20, 2020) to adjust the scatter plots in Figure 2 and 3. The average number of retweets was 3.14 and favorites was 10.20, with median for both being zero. This shows that most tweets had very less engagement with users, and certain tweets received a lot of attention (given by the maximum value of the outlier tweet in Table 1.

Table 1: Descriptive statistics for retweets and favorites per tweet

| Description | Retweets | Favorites |
|---|---|---|
| Mean | 3.142327552 | 10.20406902 |
| Median | 0 | 0 |
| Max | 7743* | 16943* |
| Min | 0 | 0 |

* Outlier tweet removed (only for figures): "ZDNet write "Researchers propose method to track coronavirus through smartphones while protecting privacy". See full original article: https://www.zdnet.com/article/researchers-invent-method-to-track-coronavirus-through-smartphones-while-protecting-our-privacy/#ftag=RSSbaffb68 All our feeds: https://secnews.physaphae.fr"

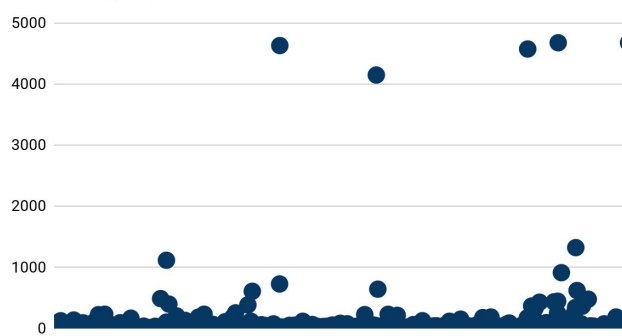
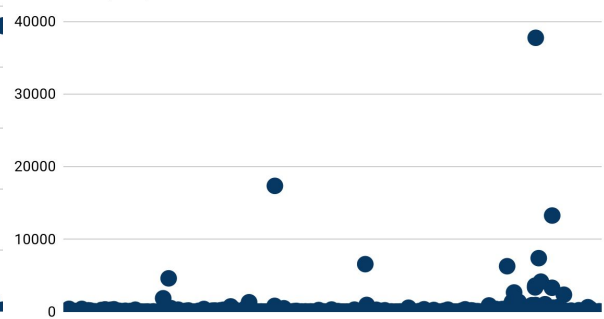

Figure 2: Retweets per tweet over time          Figure 3: Favorites per tweet over time

**Users Discussed Mostly Surveillance-Related Topics When It Comes to Privacy**
Users seem to discuss both positive and negative aspects of surveillance and loss of privacy during COVID-19. While some users are concerned about the government and private companies tracking individuals, others note the possible benefits of contact-tracing through smartphones in containing the spread. The use of devices like smartphones to collect health information and long-term surveillance in the face of European privacy laws was discussed. Similarly, users also discussed the need for surveillance to protect people. A third category of topics just inform about the situation without taking any sides like cross-country efforts in surveillance. Another common theme that emerged through topic modelling was privacy concerns about technology in use (Topic 5, 6, and 8). There seemed to be concerns regarding data security about information gathered on mobile applications. There was also discussion around the impact of data collection by companies like Google, whose products are being increasingly used for long-distance connectivity for both personal and professional reasons. Table 2 shows the ten topics modelled using LDA for both March and April along with their description.

Table 2: Latent Dirichlet Analysis of topics and matching them across March and April to find specific description.

| Topic | Description | March keywords | April keywords |
| --- | --- | --- | --- |
| 1 | Privacy concerns of tracking while staying at home during lockdown | Track, warn, covid19, the, 2020, concern, lockdown, spread, will | Privacy, news, say, time, cybersecurity, name, home, i, we |
| 2 | Public's health information tracking through smartphones | Http, news health fight, per, right, public, bluff | Health, track can, via, world, right, smartphone, issue, article, business |
| 3 | Surveillance issues of various available technology likes phones | Data, chief, start, phone, need, offer, issue | Garant, surveil, tech, get, work, privacy, individual |
| 4 | Need for surveillance to protect people in future | Pandemic, new, via, amid, people, protect, say, is, bit, us | Coronavirus, how, the, concern, covid, future, need |
| 5 | App privacy concerns in the presence of European laws like GDPR in encrypting information | Privacy, app, freedom, want, european, en, take, apr, give | Encrypt, respect, bit, law, twitter, gdpr, inform, position |
| 6 | Security of data collected (data protection) during COVID-19 | Coronavirus, surveil, security, code, How, risk, contact, digit, dataprotect, covid-19 | Covid, public, fight, contact, app, just, all |
| 7 | Worldwide surveillance at work, school to protect | Use, world, can, nation, expert, work, corona, test, what, we | Http, data, use, test, protect, person, live, school |
| 8 | Impact of and threats from technology companies like Google for privacy | Www, com, threat, covid, technology, tech, google, i | Delete, people, impact, new, want, companies, infect, go |
| 9 | Postoperative | Here, china, law, check, user, | Com, virus, case, security, |

|    | surveillance of people affected by the pandemic especially in China | twitter, operation, go | pandemic, history, possess |
|----|----|----|----|
| 10 | Contact-tracing of people in quarantine by government | Amp, lie, government, trace, first, org, now, quarantine | Will, per, state, human, response, they, talk, guest |

**Rising Negative Sentiment, but with Less People Having a Hard Opinion**

Sentiment analysis of the tweets using NLTK's Naive Bayesian classification method revealed that there were not significant changes in public sentiment over the two-month period of the novel coronavirus privacy discussions. Negative opinions about privacy concerns during COVID-19 increased from 29% to 33% from March to April 2020, with "very negative" sentiment showing a hike. Consequently, positive sentiment expressed in support of surveillance decreased over the months, going from 37% in March to 28% in April, mostly the "very positive" going down. A possible explanation of the rising negative sentiment could be due to the increasing number of news articles about security bugs in video-conferencing software like Zoom which might have led to popular frustration about privacy and security concerns about such software, which have specifically seen a hike in use due to work from home during COVID-19.

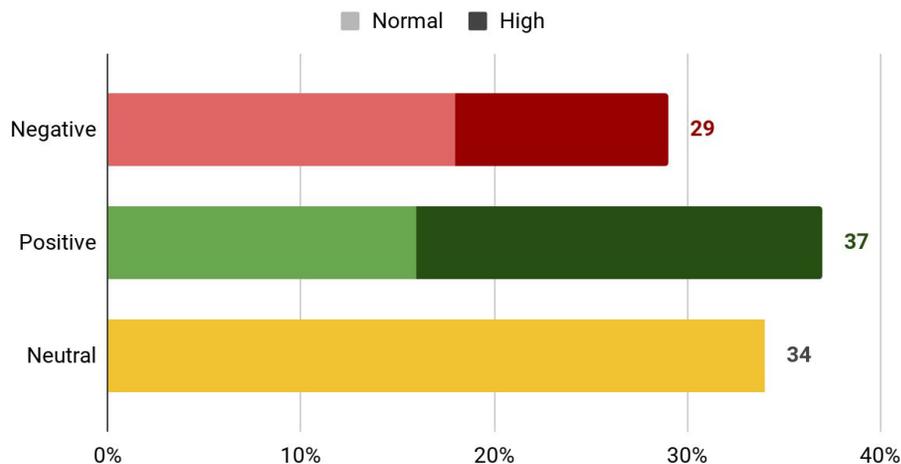

Figure 4: Higher positive sentiment about privacy and surveillance during the initial lockdown period in March

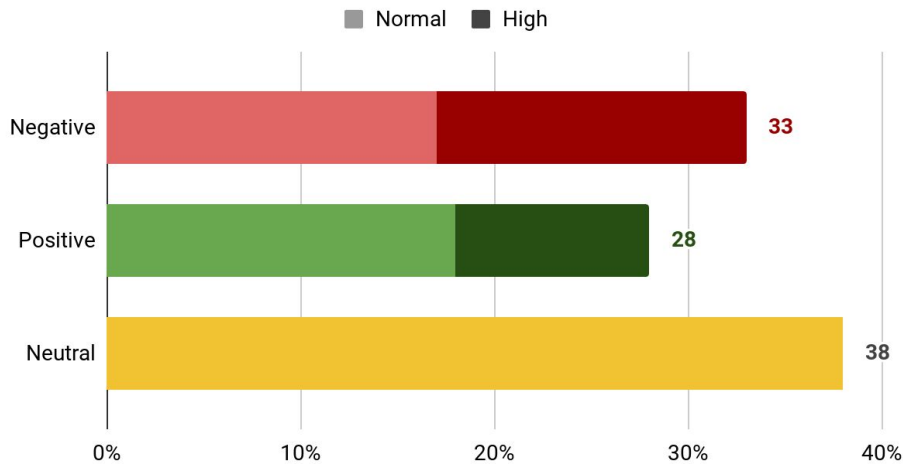

Figure 5: Higher negative sentiment about privacy and surveillance during the later lockdown period in April

However, the majority of the tweets were neutral in emotion (34% in March to 38% in April). This indicates that tweets were mostly used to report facts and information rather than expression of an opinion on privacy concerns. This is also supported by the fact that the tweet with the maximum engagement (retweets, favorites) is a short description of an article about widespread surveillance and not an expression of sentiment. Figure 4 and Figure 5 show the results of sentiment analysis for the month of March and April respectively (expressed as percentages).

## **Conclusion**

This study provides insight into privacy expectations of users on Twitter during pandemic situations when technology is used to control outbreaks. Lower privacy concerns expressed in the initial lockdown period are replaced with a more negative sentiment towards governmental and organizational efforts to maintain privacy in health information disclosure over time. More positive views towards surveillance in controlling the spread of Covid-19 with country-specific discussion topics on public and private sector efforts for technological intervention in monitoring spread of the virus is replaced by post-pandemic privacy concerns. Topic modelling indicates that the discussions about privacy is usually geared more towards surveillance, probably driven by the discussion around surveillance started by the Electronic Frontier Foundation with a COVID-19 surveillance article [3], followed by a New York Times opinion article [10]. This is supported by the hike in the number of tweets around March 23, 2020, which is the date of the published article [3]. Further research on privacy concerns of contact-tracing in order to address a public health emergency like the novel coronavirus outbreak would help form policy around long-term continuous location tracking post the pandemic in order to delete such information after use, while being instrumental in protecting public safety during the outbreak.

**Appendix**

The data and code is available here (source acknowledgement inline): https://iu.box.com/s/tzn4ak4ymyuna827cvgh5w3pyo42e7pr . Python scripts may require interpreter debugging due to the various libraries. Please use pip install to install the required libraries (especially for NLTK).